\documentclass[aps,twocolumn,floatfix]{revtex4}
\usepackage{latexsym,amssymb,graphicx,amsmath,epsfig,calc,times}

\usepackage[usenames]{color}

\citeindextrue

\def\btt1{{\tt$\backslash$\string1}}%

\def\AmS{{\protect\the\textfont2
        A\kern-.1667em\lower.5ex\hbox{M}\kern-.125emS}}

\newcommand{\kt}{k_\text{B}T}
\newcommand{\ktz}{k_\text{B}T_0}
\newcommand{\el}{\epsilon_\text{e}}
\newcommand{\ep}{\epsilon_\text{p}}
\newcommand{\conc}{c_\text{T}}
\newcommand{\ddp}{d_\text{p}}
\newcommand{\de}{d_\text{e}}
\newcommand{\phat}{\hat p}
\newcommand{\askew}{\varphi_\text{skew}}
\newcommand{\nf}{n_\text{f}}
\newcommand{\nfopt}{n_\text{f}^{*}}
\newcommand{\grod}{g_\text{rod}}
\newcommand{\gcap}{G_\text{cap}}

\newcommand{\gone}{G(1,1)}
\newcommand{\avgl}{\langle l \rangle}
\newcommand{\Gtot}{G_\text{tot}}
\newcommand{\cmc}{c_\text{csc}}
\newcommand{\mub}{\mu_\text{b}}

\begin{document}

\title{Self-limited self-assembly of chiral filaments}
\author{Yasheng Yang}
\author{Robert B. Meyer}
\author{Michael F. Hagan}
\affiliation{Department of Physics, Brandeis University, Waltham, MA, 02454}
\email{hagan@brandeis.edu}
\date{\today}
  \begin{abstract}
  The assembly of filamentous bundles with controlled diameters is common in biological systems and desirable for the development of nanomaterials. We discuss dynamical simulations and free energy calculations on patchy spheres with chiral pair interactions that spontaneously assemble into filamentous bundles. The chirality frustrates long-range crystal order by introducing twist between interacting subunits.  For some ranges of system parameters this constraint leads to bundles with a finite diameter as the equilibrium state, and in other cases frustration is relieved by the formation of defects.  While some self-limited structures can be modeled as twisted filaments arranged with local hexagonal symmetry, other structures are surprising in their complexity.
  \end{abstract}

 \maketitle

Filamentous bundles assembled from protein subunits are essential structural and regulatory components of cells and tissues. For example, filamentous actin, microtubules, and intermediate filaments assemble and disassemble to create a strong but dynamic cytoskeleton, fibrogen subunits assemble into fibrin fibers and networks to form blood clots (e.g. \cite{Weisel1987,Ryan1999}) and sickle hemoglobin assembles into fibers that impair red blood cell function (e.g. \cite{Cretegny1993,Watowich1993,Makowski1986,McDade1993,Turner2003,Madden1993}). In vivo and in vitro studies suggest that fibers with finite diameters are the stable morphology for fibrin \cite{Weisel1987,Ryan1999} under a variety of conditions, while 7-double-strand bundles of sickle hemoglobin are metastable  \cite{Turner2003,Cretegny1993,Watowich1993}, but the forces that limit filament growth remain unclear. Theoretical calculations \cite{Grason2007,Turner2003,Grason2009,Nyrkova2000,Aggeli2001} have suggested that finite bundle diameters are the thermodynamically favored state for twisted bundles assembled from chiral subunits. These calculations, however, assume specific packings of protofilaments without defects, which could relieve strain and thereby enable unbounded growth. The objective of this article is to determine, without assumptions about assembly pathways or assemblage geometries, if chirality can result in stable bundles with finite diameters. We construct a model subunit with simple pairwise chiral interactions that drive assembly into filamentous bundles, and combine umbrella sampling \cite{Frenkel2002} and forward flux sampling \cite{Allen2005} to explore the structures that spontaneously assemble for varying degrees of chirality. The simulations demonstrate that chirality can result in regular self-limited bundles for a range of interaction strengths, but that stronger interactions enable defects which give rise to branched networks or irregular bundles.

Drawing conclusions about self-limited growth in a macroscopic system from simulations with a finite number of subunits is challenging--one must distinguish between simulations in which growth terminates due to physical constraints from those in which the system runs out of subunits \cite{Huisman2008}. To overcome this limitation, we simulate the grand canonical ensemble ($\mu V T$), in which growth cannot terminate because of subunit depletion, since the system is coupled to an unlimited bath of free subunits. Before discussing the simulations, we consider the conditions for self-limited filamentous assembly at fixed total subunit concentration (the $NVT$ ensemble).

 We build on the theory for cylindrical micelles \cite{Safran1994, Ben-Shaul1985,Gelbart1996} to write the free energy for a twisted bundle comprised of $\nf$ filaments and $n$ subunits as $G(n,\nf)=n (\grod(\nf)-\gone) + \gcap(n,\nf)$ with $\grod$ the subunit chemical potential in the cylindrical `body' of the bundle and $\gcap$ the excess chemical potential of the subunits in the `caps' at either end. $\gcap$ and $\grod$ are independent of length for long bundles, but depend on the bundle diameter, or $\nf$, since subunits at different radii experience different environments. For a solution with fixed total subunit density $\conc$, minimizing the solution free energy density $\Gtot=\sum_{n,\nf} \rho(n,\nf) G(n,\nf) -TS$ with the mixing entropy $S=-k_\text{B} \sum_{n,\nf} \rho_{n,\nf} \ln (\rho(n,\nf) \sigma^3)$ with $\sigma$ the subunit size gives the  law of mass action result for equilibrium bundle densities\begin{align}
\rho(n,\nf)=\exp[\beta(\mu n - G(n,\nf))]
\label{eq:one}
\end{align}
with $\beta=1/\kt$ and the monomer chemical potential $\mu = \gone+\kt \ln( \rho(1,1) \sigma^3)$.  The onset of spontaneous assembly is identified with the concentration at which half of the subunits are assembled into bundles with the other half free in solution, which is given by by \cite{Gelbart1996} $\cmc\approx \exp[\beta(\grod-\gone)]$. By minimizing $\Gtot$ with respect to $n$ and $\nf$ we obtain that bundles follow an exponential distribution of mass-averaged lengths $P(l)\sim l  e^{-l/\avgl}$, with $l=n/\nf$ and the mean length $\avgl\simeq[\conc \exp(\gcap)]^{1/2}$. However, if there is a bundle diameter ($\nfopt$) that minimizes $\grod$, the distribution will be sharply peaked about $\nfopt$ when $\avgl\gg1$, and growth will be self-limited. Thus, we begin by measuring $\grod$ as a function of $\nf$.

We note that in Ref.~\cite{Huisman2008} filamentous bundle formation is described in terms of the theory of linear polymerization \cite{Douglas2009,Sciortino2007} and bundling of linear polymers. That analysis would be complicated for our model since the subunit binding free energy depends crucially on the number of bundled filaments due to the twist imposed by chirality.

\begin{figure}
\epsfig{file=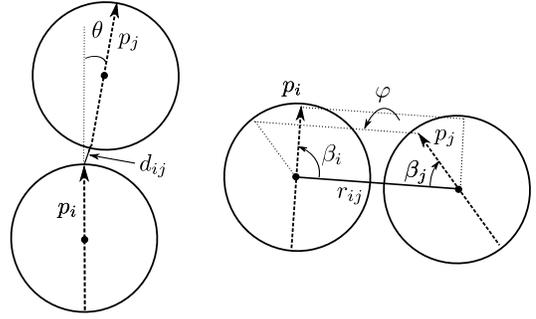,width=.8\columnwidth}
\caption{\label{fig:equitorial} Subunit model.
{\bf (left)} The pole-pole interaction. $d_{ij}$ is the distance between the corresponding poles, and $\theta$ is the angle
between $\hat p_i$ and $\hat p_j$.
{\bf (right)} The equatorial interaction. $\varphi$ is the dihedral angle
between the plane $(\vec r_{ij},\hat p_i)$ and the plane $(\vec r_{ij},\hat p_j)$. For all simulations reported in this work, the polar interaction strength is $\ep=14\ktz$, the lateral interaction strength is $\el=2.95\ktz$, and the distance and angle tolerance parameters are $\ddp=0.1\sigma$, $\de=0.1\sigma$, $\beta_\text{max}=1$, $\phi_\text{max}=\theta_\text{max}=0.25$ with angles in radians. The temperature is $T=T_0$, except for the simulation of $\askew=0$, for which $T=1.05T_0$. The GC bath chemical potential is $\mub=\kt\ln 0.01$. For this model $\gone=\kt \ln 8\pi^2$ due to rotational entropy.
}\end{figure}

{\bf Subunit Model.}
We consider spherical subunits of diameter $\sigma$ that are endowed with a polar orientation unit vector $\phat$ and have azimuthal symmetry, ( Fig.~\ref{fig:equitorial}). Subunits have pairwise interactions that drive north-to-south pole alignment into filaments and weak equator-to-equator interactions that drive bundling of filaments. The interaction between two subunits $i$ and $j$ is given by
\begin{align}
\lefteqn{V(i,j) = V_\text{h}(r_{ij}) + V_\text{p}(\cos\theta,d_{ij})} \nonumber \\
& &	+ V_\text{p}(\cos\theta, d_{ji})
	+ V_\text{e}(\vec r_{ij}, \hat p_i, \hat p_j)
\label{eq:Vij}
\end{align}
where $\vec r_{ij}=\vec r_j - \vec r_i$ is the interparticle displacement, $r_{ij}=|\vec r_{ij}|$,
$\cos \theta=\hat p_i \cdot \hat p_j$ gives the angle between the two polar directions, and
$d_{ij}=|(\vec r_j - \sigma \hat p_j/2) - (\vec r_i + \sigma \hat p_i/2)|$
is the distance between poles. Excluded volume is imposed by a hard-sphere interaction
\begin{align}
V_\text{h}(r)=\left\{\begin{matrix}\infty& r<\sigma\\0 & r\geq \sigma\end{matrix}\right.
\label{eq:Vh}.
\end{align}
The pole-pole interaction is given by
\begin{align}
V_\text{p}(\cos\theta,d)=-\ep H(\ddp - d)f(\theta, \theta_{\text{max}})
\end{align}
with $H(x)$ the Heaviside step function, $\ep$ the pole-pole interaction strength,
$\ddp$ the distance tolerance, and $\theta_{\text{max}}$ the angle tolerance. Parallel alignment is driven by
\begin{align}
f(y, y_\text{max})&=\left\{\begin{matrix}\exp(-y^2/y_\text{max}^2) & y < y_\text{max} \\
		                0 & y \geq y_\text{max}  \end{matrix}\right.
\end{align}
The equatorial interaction is given by
\begin{align}
\lefteqn{V_\text{e}(\vec r_{ij}, \hat p_i, \hat p_j) =
- \el H(\de - r_{ij})H(\beta_\text{max} - \beta_{ij}) } \nonumber \\
& &H(\beta_\text{max}-\beta_{ji})f(\varphi - \askew, \varphi_\text{max})
\label{eq:Ve}
\end{align}
where $\el$ is the equatorial interaction strength, $\de$ is the equatorial distance tolerance, and the width of the equatorial interaction band is set by $\beta_\text{max}$ with $\cos\beta_{ij}=|\vec r_{ij}\cdot \hat p_i|/r_{ij}$.
The final factor in Eq.~\ref{eq:Ve} measures the degree of twist, with $\varphi$ as the dihedral angle
between the plane $(\vec r_{ij},\hat p_i)$ and the plane $(\vec r_{ij},\hat p_j)$ (Fig.~\ref{fig:equitorial}), which can be calculated from the following relations, with $\vec q_{ij} = \vec r_{ij} \times \hat p_i$
\begin{align}
\sin\varphi&=\frac{\left(\vec q_{ij}\times \vec q_{ji}\right)\cdot \vec r_{ij}}{|\vec q_{ij}||\vec q_{ji}| |r_{ij}|}, \quad
\cos\varphi=\frac{\vec q_{ij}\cdot \vec q_{ji}}{|\vec q_{ij}||\vec q_{ji}|} 
\label{eq:alpha}
\end{align}
 The degree of chirality is dictated by the preferred skew angle $\askew$, with $\varphi_\text{max}$ as the tolerance for deviations from the preferred skew.

{\bf Simulations.}
We explore assembly with Monte Carlo, with random translations and rigid body rotations of subunits accepted according to the Metropolis criterion\cite{Frenkel2002}. We focus on parameters for which nucleation is a rare event and bundles grow by addition of monomers, and thus we use only single particle moves. To ensure that self-limited growth is not a result of subunit depletion, we sample the grand canonical ensemble by coupling the system to a subunit bath at chemical potential $\mub$ with subunit insertion/deletion moves \cite{Frenkel2002}. Since there are large nucleation barriers, we employ umbrella sampling simulations\cite{Frenkel2002} to calculate the free energy. We use the total bundle size $n$ as the reaction coordinate, and measure the probability $p(n,\nf)$ that a particular subunit is in a bundle of size $n$, using a series of windows in which hard walls constrain the size of that cluster to a range of $n$. The free energy is then obtained from Eq.~\ref{eq:one} with $G(n,\nf) = - \kt \ln [\rho(n,\nf) \sigma^3] + \mub$, with $\rho(n,\nf)=p(n,\nf)/n$ \cite{Maibaum2008,Wolde1998}.

\begin{figure}
\epsfxsize=0.59\columnwidth\epsfbox{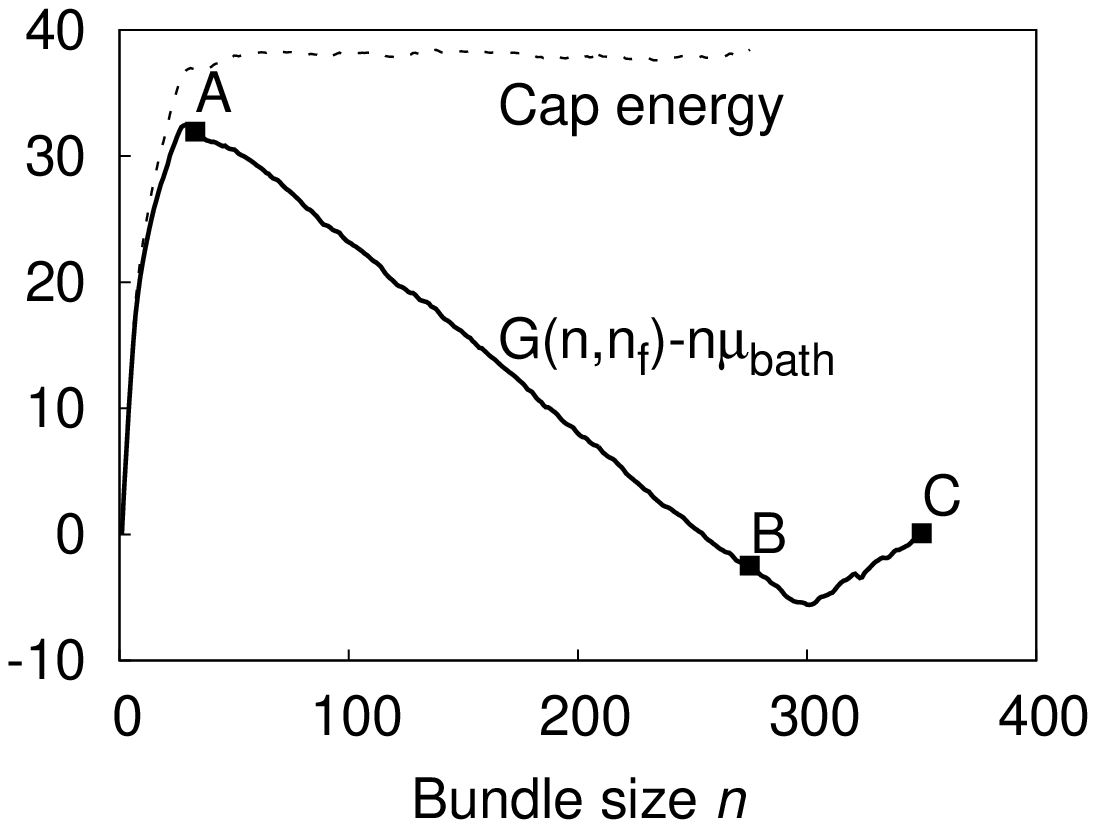}
\epsfxsize=0.39\columnwidth\epsfbox{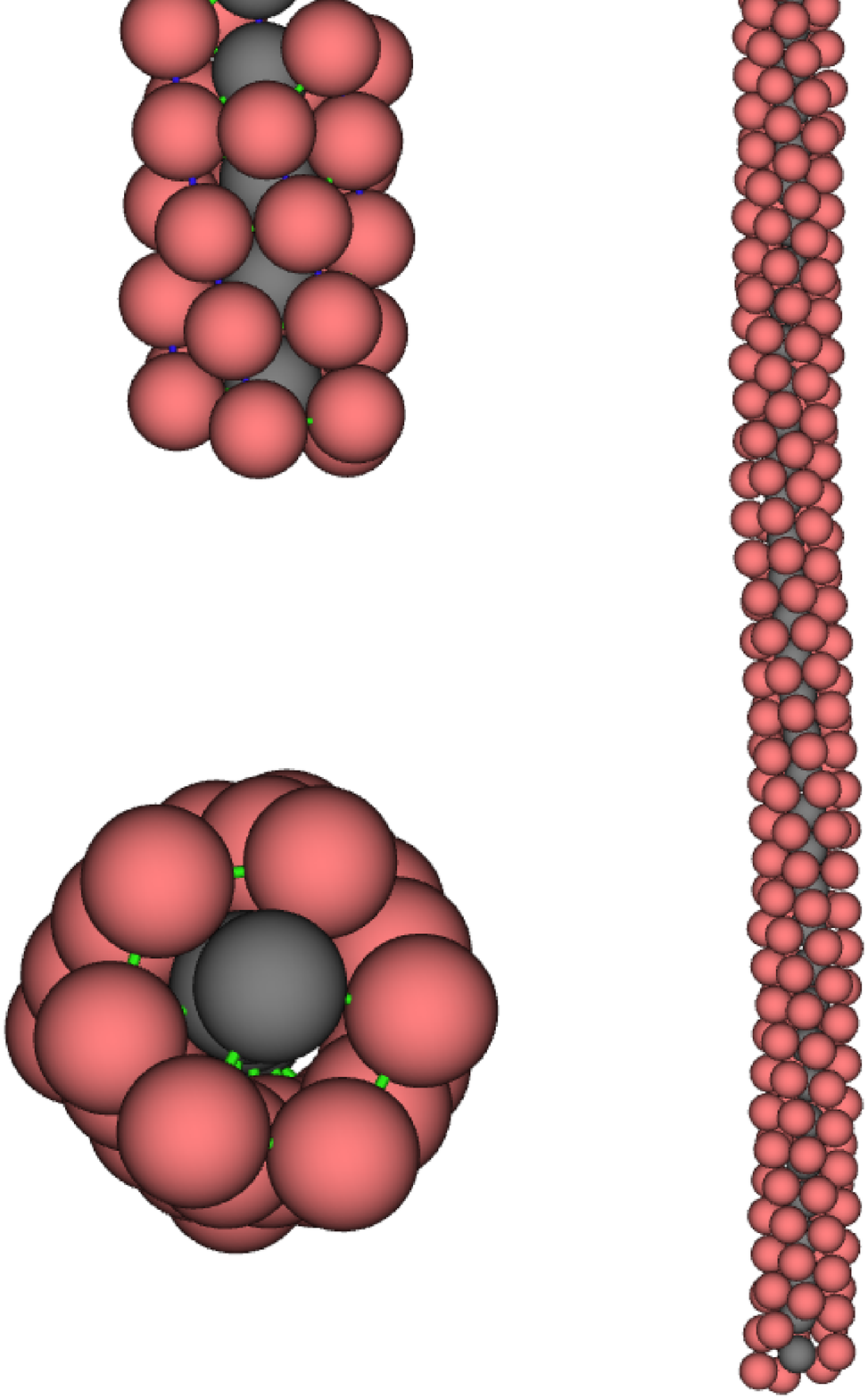}
\caption{\label{fig:freeEnergySkew}.
{\bf (left)} Free energy $G$ and cap free energy $\gcap$ as functions of the number of subunits $n$ with $\askew=0.38$ obtained from umbrella sampling. The rise in free energy at large $n$ occurs when the bundle reaches the hard system boundary and begins to grow a third layer. {\bf (right)} Structures corresponding to the indicated points on the free energy plot.
}
\end{figure}

The free energy for $\askew=0.38$ and snapshots of representative bundle configurations are shown in Fig.~\ref{fig:freeEnergySkew}.  We see a rapid rise in free energy at small $n$ during which short structures with $\nf=3$, 4 , and 5 filaments appear successively, followed by the critical nucleus with $\nf=7$ and $n \approx 25$ subunits (Fig.~\ref{fig:freeEnergySkew}A. The free energy is unfavorable below this size because the majority of subunits have unsatisfied lateral and/or polar contacts (see Cap Free Energy below). After reaching the critical nucleus, the bundle grows lengthwise in both directions, while maintaining the same structure, and the free energy decreases linearly. As shown below, the $\nf=7$ structure corresponds to the optimal bundle structure for $\askew=0.38$ and thus further lateral growth is unfavorable. The slope of the free energy in this region corresponds to the chemical potential $\grod(\nf=7)$.

While the exponential distribution of filament lengths is derived above for the NVT ensemble, in the $\mu$VT ensemble the bundle will continue to grow lengthwise indefinitely (provided that $\grod<-\mub$). To evaluate the free energy of lateral growth, we imposed hard spherical boundary conditions with a diameter $D=44\sigma$, which is large enough that bundle properties are independent of $D$. Upon reaching the boundary, the bundle grows by increasing its diameter, which results in the increasing free energy at large $n$ in Fig.~\ref{fig:freeEnergySkew}.

{\bf Cap free energy.}
 The cap free energy is calculated using $\gcap(n,\nf) = G(n,\nf)-n(\grod(\nf)-g_1)$, with $\grod$ obtained from the slope of the linear regime in the free energy (or from Fig.~\ref{fig:chemical-potential} below). As shown in Fig.~\ref{fig:freeEnergySkew}, $\gcap$ rises rapidly until saturating at the critical nucleus with $\gcap\approx 37 \kt$ for $\askew=0.38$. This value is similar to cap energies measured for cylindrical micelles \cite{Gelbart1996} and corresponds to a large average bundle length in the canonical ensemble; using the measured $\grod$ and $\gcap(n,\nf)$ we solved  Eq.~\ref{eq:one} to obtain a mass-averaged bundle size of about $10^8$ subunits at the CSC.
 Although the magnitude of $\gcap$ depends on the strength of the polar bonds ($\ep=14 \kt$), we find that robust bundle formation requires $\ep \gg \el$, implying that large cap free energies and hence large average filament lengths are general.

Similar results were obtained for lower skew angles $\askew=0.32$ and $\askew=0.25$  at low $n$, but convergence in the free energy calculation was questionable because transitions between different values of $\nf$ were rare at large $n$. We overcame this limitation as follows.

{\bf Self-limited bundle diameters depend on preferred skew.}
In the umbrella sampling simulations the system adopts the number of filaments $\nfopt$ that minimizes $G$ for a given $n$. To determine the dependence of the subunit free energy $\grod$ on $\nf$,  we performed additional sets of `constant filament number' umbrella sampling (CFNUS) simulations in which $\nf$ and $n$ are constrained. A small structure with $\nf$ filaments is extracted from an umbrella sampling simulation, and subjected to a simulation in which any move that changes $\nf$ is rejected. Specifically, moves which cause the number of subunits in any filament to differ by more than 3 are rejected and the mean of each filament along the bundle axis must remain within $2\sigma$ of the bundle center.  These additional requirements constrain the configurations of the cap and hence affect $\gcap$, but do not affect $\grod$ in long bundles, which is determined from $\grod(\nf) = (\partial G(n,\nf)/\partial n)_{\nf}$ at large $n$ (when $\grod$ becomes independent of $n$). This procedure is repeated for all commonly observed morphologies with a given $\nf$.

\begin{figure}
\epsfxsize=0.61\columnwidth\epsfbox{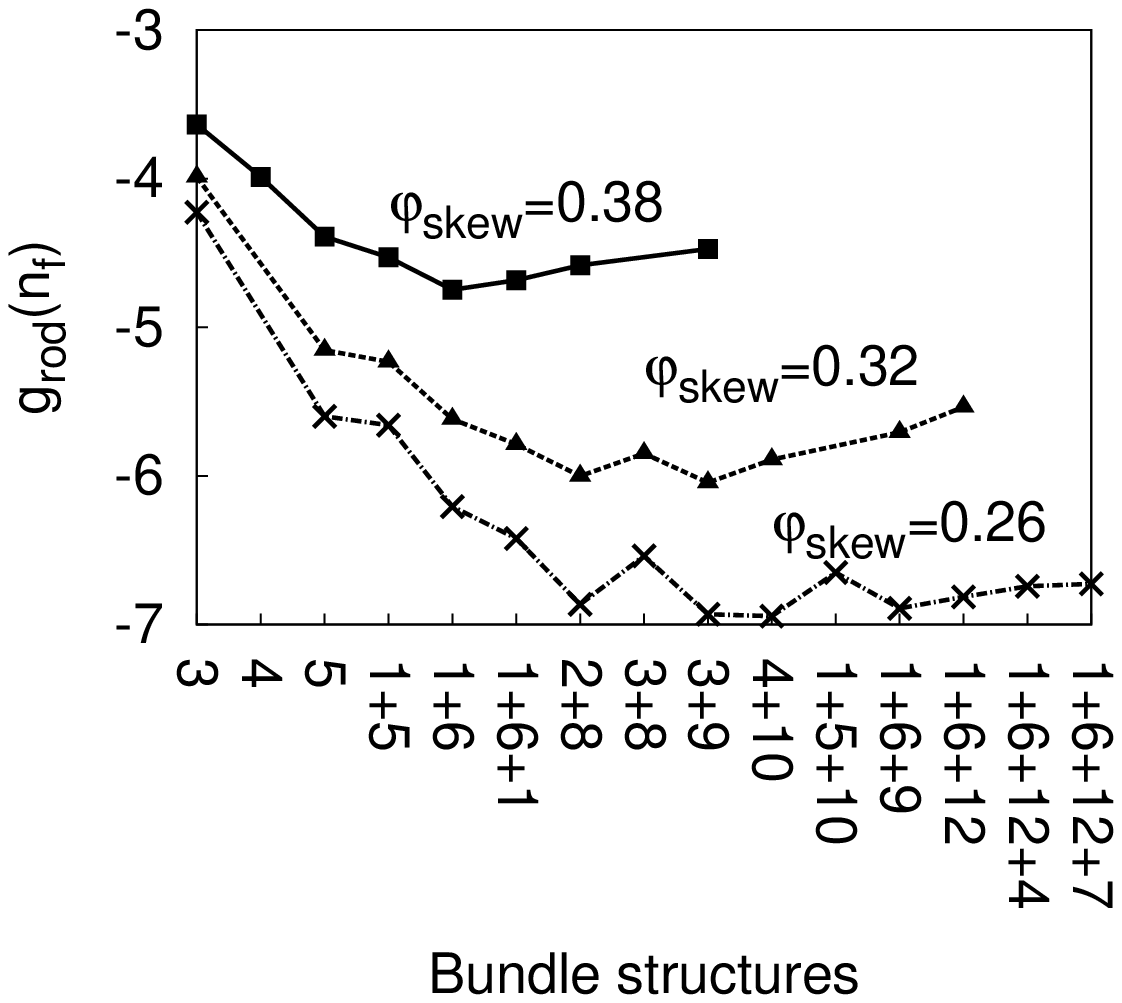}
\epsfxsize=0.37\columnwidth\epsfbox{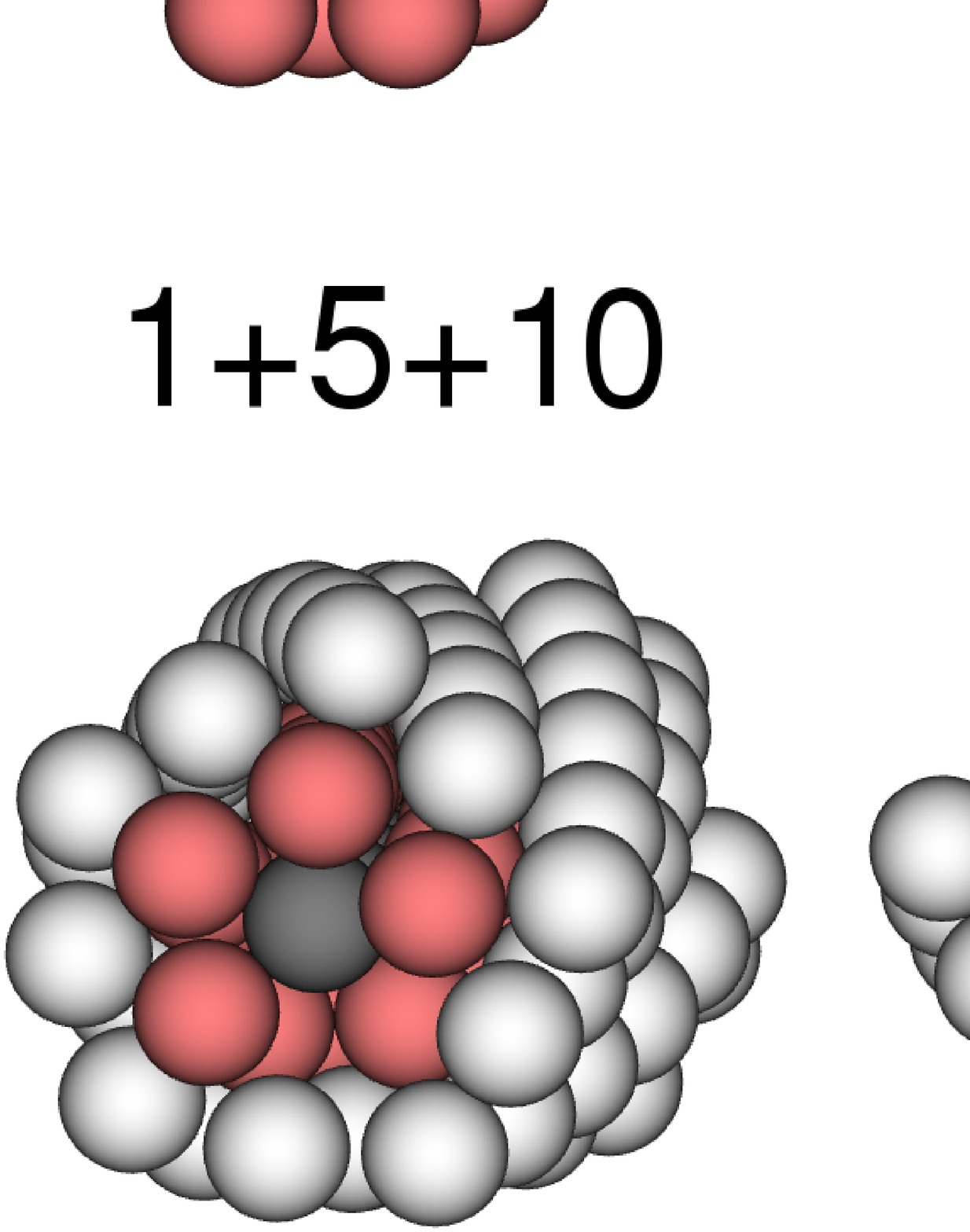}
\caption{{\bf (left)} Subunit free energy as a function of filament structure for three skew angles from the constrained umbrella sampling (CFNUS). Structures are labeled with the number of filaments in each layer starting from the center. Results are shown only for the lowest free energy morphology at each value of $\nf$. {\bf (right)} Snapshots from CFNUS simulations, shown in cross-section view, illustrate optimal bundle morphologies for some values of $\nf$.
 \label{fig:chemical-potential}
}
\end{figure}

The chemical potential $\grod$ calculated from the CFNUS simulations is shown as a function of bundle size for three preferred skew angles in Fig.~\ref{fig:chemical-potential}. In each case, there is an optimal bundle diameter, or number of filaments $\nfopt$. Although the minima appear shallow, the large average bundle lengths calculated above ensure that the free energies for different bundle morphologies differ by many $\kt$. Note that $\nfopt=7$ for $\askew=0.38$, in agreement with the unrestrained umbrella sampling.  Furthermore, the slope of the free energy in the linear region of  Fig.~\ref{fig:freeEnergySkew} gives $\grod=-4.75$ (via Eq.~\ref{eq:one}), which matches the chemical potential determined for $\nf=7$ in Fig.~\ref{fig:chemical-potential}, showing that the two protocols agree.

\begin{figure}
\epsfxsize=0.4\columnwidth\epsfbox{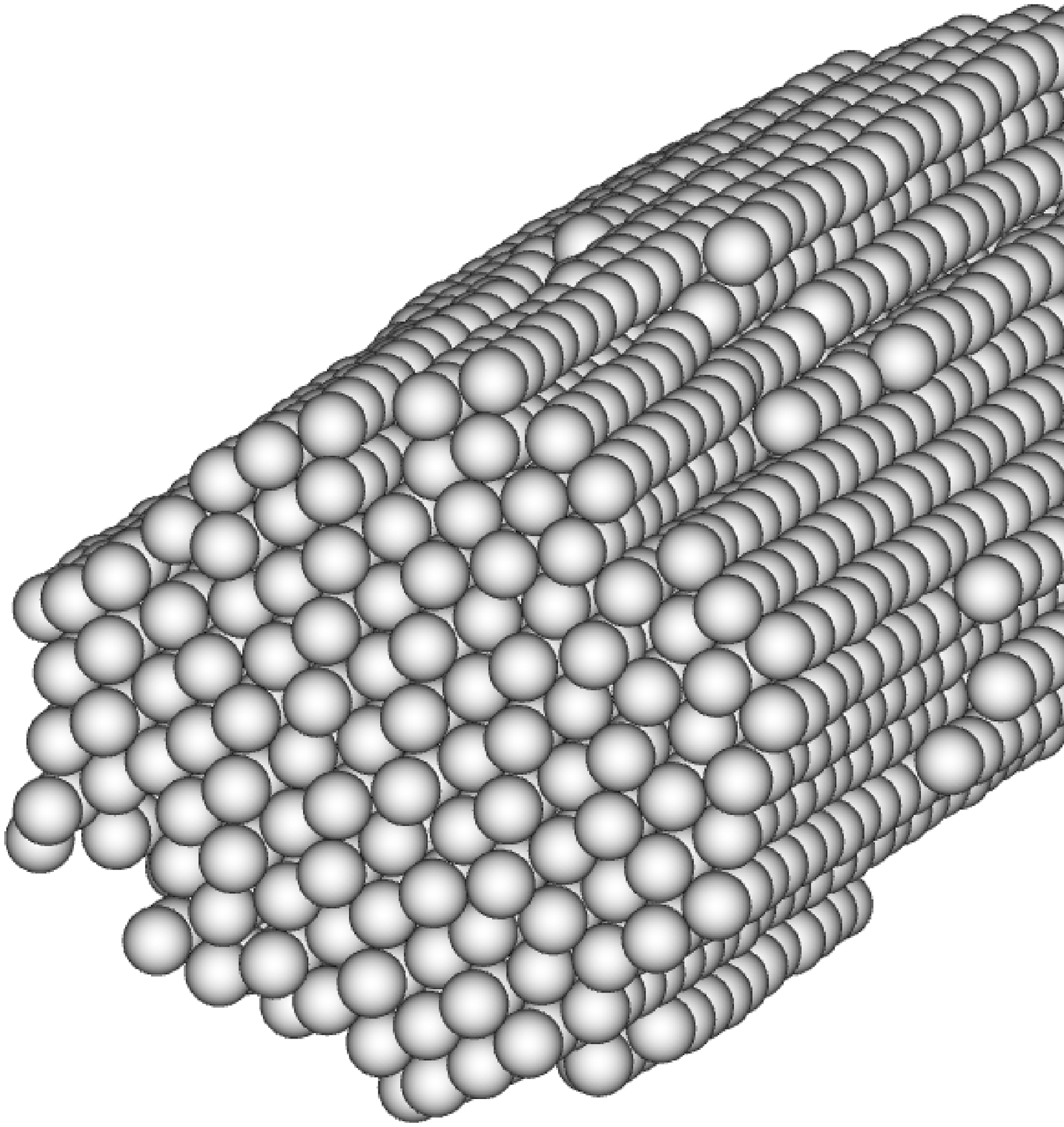}
\epsfxsize=0.58\columnwidth\epsfbox{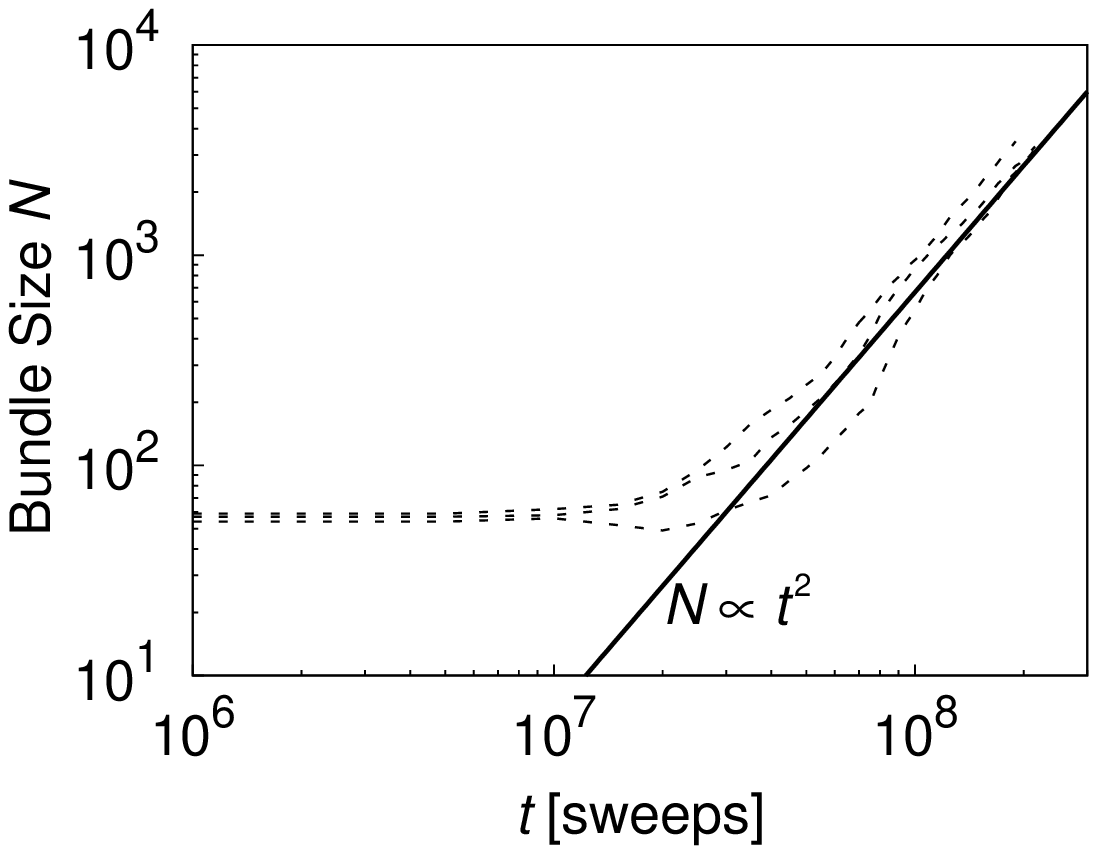}
\caption{ {\bf (a)} Bundle size as a function of Monte Carlo (MC) sweeps for three trajectories with $\askew=0$ and $T=1.05T_0$. Trajectories were initiated using forward flux sampling (FFS) as described in the text, dynamics are shown after FFS ended. The solid line indicates the scaling $n(t) \sim t^2$. {\bf (b)} A snapshot from one trajectory in {\bf (a)}.
 \label{fig:traj0structure}
}
\end{figure}

The existence of an optimal diameter can be understood as follows. Adding layers decreases energy, since subunits in the outermost layer have unsatisfied lateral contacts. However, the preferred skew $\askew$ causes filaments to tilt with respect to the central filament; tilt increases with layer number for preferred skew angles $\askew$ above a critical value.  Due to tilt, filaments trace a curved path around the bundle, which requires unfavorable bending of polar bonds. Furthermore, complete formation of lateral bonds between layers, requires that the exterior filaments stretch while the interior filaments compress. With each additional layer, the degree of extension and compression increases. These effects overwhelm the energetic benefit of adding an additional layer at the optimal bundle diameter $\nfopt$. Since the magnitude of the tilt increases with $\askew$, $\nfopt$ decreases with increasing $\askew$ (Fig.~\ref{fig:chemical-potential}).  In agreement with this explanation, the chemical potential for $\askew=0$ (corresponding to achiral interactions) does NOT show a minimum, and as shown below (Fig.~\ref{fig:traj0structure}b) bundles with $\askew=0$ have unbounded lateral growth. This observation confirms that chirality is the reason for self-limited bundle sizes in this model.

Interestingly, the optimal bundle morphology for a given number of filaments changes with $\nf$. As shown in  Fig.~\ref{fig:chemical-potential}b, the central layer of the bundle can vary between 1 and 4 protofilaments, and usually corresponds to the structure that maximizes rotational symmetry. While these are the lowest free energy structures at each value of $\nf$ among those taken from unrestrained umbrella simulations, we cannot rule out lower free energy morphologies that we did not test.

{\bf Dynamics.}
Having shown that self-limited bundles are the equilibrium state for our model chiral subunits, we now demonstrate that they are also the kinetically selected state. Since bundle formation is not accessible by straightforward dynamical simulations due to the large nucleation barrier (Fig.~\ref{fig:freeEnergySkew}), we used forward flux sampling (FFS) \cite{Allen2005} to obtain an unbiased ensemble of assembly trajectories.
  We used the bundle size $n$ as the order parameter and performed FFS until bundles reached a size larger than the critical nucleus (it is not necessary that the order parameter be a good reaction coordinate, although a bad order parameter can inhibit convergence). At this point, FFS was no longer needed, and we continued the simulations with straightforward dynamic Monte Carlo; nucleated bundles readily grow in length until they reached the imposed boundary ($D=32\sigma$), but lateral growth terminates at $\nf=7$, in agreement with the equilibrium calculations. The dynamic nucleation pathways observed with FFS closely follow the minimum free energy pathway observed with umbrella sampling (Fig.~\ref{fig:freeEnergySkew}b), indicating that structures below the nucleus size achieve relative equilibrium quickly in comparison to the nucleation time \cite{Endres2002,Hagan2010}.

In contrast, assembly trajectories for $\askew=0$ (Fig.~\ref{fig:traj0structure}) demonstrate unbounded lateral growth, and after reaching a size of $n\approx 100$ scale as $n(t) \sim t^2$, with $t$ the number of Monte Carlo sweeps.  This scaling is consistent with growth dominated by subunit addition to the bundle body, and $\grod$ independent of $\nf$, in agreement with the free energy calculations. Although the bundle grows with hexagonal order, we note that pentagonal defects become trapped within the assemblage, particularly during rapid growth that occurs under stronger interactions.

\begin{figure}
\epsfxsize=0.5\columnwidth\epsfbox{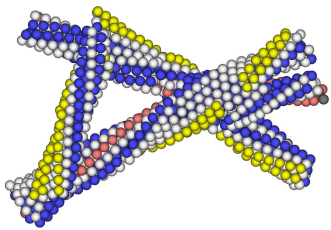}
\caption{A branched bundle formed in a dynamical trajectory with $\askew=0.32$. \label{fig:branchedSnapshot}}
\end{figure}

{\bf Strong interactions lead to branched networks.}  Bundles with $\nf>\nfopt$ tend to form defects that relieve strain. In some cases these defects serve as nucleation points for branching, which enables further growth until the branch reaches its optimal diameter and a system boundary. As shown in Fig.~\ref{fig:branchedSnapshot}, each branch maintains a finite radius. Additional branching and bundle growth lead to a branched network (see Ref.~\cite{Douglas2009} for further discussion of branched bundles). We also observe branching in the canonical ensemble simulations with strong interactions. While this simple model is not intended to represent a particular molecule and crowding affects assembly at high density \cite{Madden1993}, we note that fibrin clots are composed of branched networks of fibrin bundles with uniform bundle radii (e.g. \cite{Ryan1999}).

In conclusion, we simulated the assembly of subunits with a simple potential that drives the formation of filamentous bundles. For moderate interaction strengths, the local packing constraints that arise due to chirality cause assembly to terminate at a finite bundle diameter, while bundles propagate easily in length. Stronger interactions, however, lead to defects which enable the formation of multiply connected branched networks. The optimal morphologies of assembled bundles with different numbers of filaments have different symmetries. The simulation results indicate that spontaneously assembled structures can deviate significantly from regular hexagonal bundles, and thus it is important to evaluate assembly behavior with dynamical algorithms that do not impose particular assembly pathways or morphologies. The approach we have adopted to evaluate self-limited growth in a finite-sized simulation could be used to understand specific biological molecules; for example, a patchy-sphere model could be constructed from atomic-resolution structures of sickle hemoglobin in order to understand the effects of chirality and sphere-packing on hemoglobin filament assembly \cite{Makowski1986,McDade1993,Turner2003}.

\bibliographystyle{apsrev}

\end{document}